\documentclass[nonacm, sigconf]{acmart}

\renewcommand\footnotetextcopyrightpermission[1]{} 

\usepackage{graphicx}
\usepackage{subcaption}
\usepackage{float} 
\usepackage{afterpage}
\usepackage{bm}
\usepackage{colortbl}
\usepackage{titlesec}
\usepackage{multirow}

\definecolor{lightgray}{cmyk}{0, 0, 0, 0.173}

\AtBeginDocument{%
  }

\begin{document}

\title{Popular News Always Compete for the User’s Attention!\\ \textit{POPK}: Mitigating Popularity Bias via a Temporal-Counterfactual}

\author{Igor L.R. Azevedo}
\orcid{0000-0001-5144-825X}
\affiliation{%
  \institution{The University of Tokyo}
  \city{Tokyo}
  \country{Japan}
}
\email{azevedo-igor@g.ecc.u-tokyo.ac.jp}

\author{Toyotaro Suzumura}
\orcid{0000-0001-6412-8386}
\affiliation{%
  \institution{The University of Tokyo}
  \city{Tokyo}
  \country{Japan}
}
\email{suzumura@acm.org}

\author{Yuichiro Yasui}
\orcid{0000-0002-4175-9318}
\affiliation{%
  \institution{Nikkei Inc.}
  \city{Tokyo}
  \country{Japan}
}
\email{yuichiro.yasui@nex.nikkei.com}


\begin{abstract}
  In news recommendation systems, reducing popularity bias is essential for delivering accurate and diverse recommendations. This paper presents \textit{POPK}, a new method that uses temporal-counterfactual analysis to mitigate the influence of popular news articles. By asking, \textit{"What if, at a given time $t$, a set of popular news articles were competing for the user's attention to be clicked?"}, \textit{POPK} aims to improve recommendation accuracy and diversity. We tested \textit{POPK} on three different language datasets (Japanese, English, and Norwegian) and found that it successfully enhances traditional methods. \textit{POPK} offers flexibility for customization to enhance accuracy and diversity, alongside providing distinct ways of measuring popularity. We argue that popular news articles always compete for attention, even if they are not explicitly present in the user's impression list. \textit{POPK} systematically eliminates the implicit influence of popular news articles during each training step. We combine counterfactual reasoning with a temporal approach to adjust the negative sample space, refining understanding of user interests. Our findings underscore how \textit{POPK} effectively enhances the accuracy and diversity of recommended articles while also tailoring the approach to specific needs.
\end{abstract}



\maketitle

\section{Introduction}
In today's rapidly evolving online news landscape, recommender systems play a crucial role in tailoring content to meet the unique interests and preferences of each user. Over time, significant efforts have been made to refine these systems, with a focus on delivering more personalized and relevant news recommendations \cite{10.1145/3269206.3269307, qi2021personalized, qi-etal-2021-uni-fedrec, 10.5555/3491440.3491858, yi-etal-2021-efficient, 10.1609/aaai.v33i01.33015973, wu2019neural, wu-etal-2019-neural-news, an-etal-2019-neural}.

For instance, NRMS \cite{wu-etal-2019-neural-news} learns representations of users from their browsed news and uses multihead self-attention to capture the relatedness between the news. Similarly, NAML \cite{wu2019neural} employs an attentive multi-view learning model to learn unified news representations from titles, bodies, and topic categories. In the user encoder, they learn the representations of users based on their browsed news and apply an attention mechanism to select informative news for user representation learning. In a similar direction, LSTUR \cite{an-etal-2019-neural} propose a neural news recommendation approach which can learn both long- and short-term user representations. 

More recently, the GLORY model \cite{Yang_2023} integrates a global graph with local user interactions, offering a comprehensive understanding of user preferences through its Global-aware Historical News Encoder. Addressing the challenge of aligning candidate news with user interests, the CAUM model utilizes a candidate-aware self-attention network \cite{qi2022news}. Together, these models highlight ongoing efforts to refine techniques that enhance the personalization and precision of news recommendations. LANCER \cite{Bae_Ahn_Lee_Kim_2023} introduces an approach that leverages the concept of news lifetime to incorporate temporal relevance. This method generates the negative sample space of the model by recognizing that news articles influence users only for a limited period. 

Despite efforts to better tailor content to user needs, an increasingly significant issue is popularity bias, where the prominence of popular articles tends to dominate recommendations, potentially narrowing the range of viewpoints available to users. This bias can confine users to content that closely aligns with their established preferences or sentiments, rather than exposing them to a broader spectrum of news topics and perspectives. For example, SentiRec \cite{wu-etal-2020-sentirec} introduces a sentiment diversity-aware neural news recommendation approach aimed at diversifying recommended news articles based on sentiment. In another approach, recent strategies have explored methods like MANNeR \cite{iana2024train}, which prioritizes optimizing recommendation diversity alongside performance metrics, aiming for balanced evaluations across multiple criteria. Additionally, the PP-Rec model \cite{qi2021pprec} addresses the cold-start problem by incorporating ranking scores for recommending candidate news. It also accounts for the time-aware popularity of candidate news to eliminate popularity bias in user behaviors, thereby enabling more accurate interest modeling.

As highlighted by \cite{Wei_2021}, there exists a notable bias towards recommending more popular items in training data, leading to an over-representation of these items in recommendations. This phenomenon indicates a tendency for models to prioritize popular items over personalized user-item matches. This bias stems from the training process, where the model's objective often incentivizes recommending popular items more frequently to minimize loss, thereby adjusting parameters in favor of such recommendations. However, this approach undermines the accuracy of understanding user preferences and diminishes the diversity of recommended items. Moreover, this popularity bias can exacerbate the Matthew Effect \cite{Perc_2014}, where already popular items receive more recommendations, further amplifying their popularity.

Considering these challenges, we aimed to enhance existing methods to improve accuracy while addressing the impact of popularity in news recommendations. We began by examining how users typically access news portals. It is uncommon for users to rely solely on one platform, as they are influenced by various media sources such as social networks, radio, and online news portals. Consequently, even on platforms that tailor content based on a user's past behavior, \textit{external information} influences their decision to click on a news article. Each time a user accesses a news platform, they carry knowledge of current events from other sources. However, current recommender models do not account for this external information. To the best of our knowledge, they overlook the \textbf{implicit} influence of popular news articles on user decisions in a temporal context. We propose a method to eliminate the \textbf{implicit} influence and \textbf{explicitly} incorporate it into the negative sampling process with a temporal-counterfactual approach. This enhancement allows the model to gain a better understanding of the user's true preferences.

Our method \textit{POPK} has led to noticeable enhancements in the performance metrics linked with existing deep learning based recommender models. Our contributions can be outlined as follows: \textbf{(1)} A straightforward yet efficient strategy to mitigate the impact of popular news articles, subsequently improving existing methods in terms of diversity and accuracy; \textbf{(2)} Demonstrations highlighting that \textit{POPK} not only significantly enhances the performance of leading models across various metrics but also offers flexibility for easy adaptation to individual cases and seamlessly integration; \textbf{(3)} Thorough experiments conducted on real-world data to validate the efficacy of our proposed model.

\section{Related Work}

\subsection{Popularity bias in News Recommendations}

This study examines the influence of popular items in news recommender systems, a topic well-established in the literature \cite{qi2021pprec, 10.1145/3583780.3615272, 10.1145/3109859.3109912, 10.1145/3077136.3080836, 10.1145/3209978.3210014, he2017neural, 10.1007/s11257-015-9165-3, 10.1145/3564284, 9338389, zheng2021disentangling}. For example, CR Framework \cite{10.1145/3404835.3462962} constructs a causal graph to depict essential cause-and-effect relationships in the recommendation process. Through multi-task learning during training, they assess the impact of each causal factor, while during testing, they employ counterfactual inference to mitigate the influence of item popularity. This method introduces a versatile enhancement applicable across various models, seamlessly integrating into existing frameworks. Similarly, MACR \cite{Wei_2021} adopts a causal graph approach to analyze recommendation models, addressing issues like clickbait through counterfactual inference. By simulating a scenario where items are evaluated solely on their exposure features visible to users before making click decisions, they gauge user click likelihood, thereby mitigating the direct impact of exposure features and mitigating clickbait.

Similar to approaches found in \cite{10.1145/3404835.3462962, Wei_2021}, our method also employs a counterfactual reasoning to improve recommendation accuracy and mitigate popularity bias. However, our approach raises a fundamental question based on the temporal nature of popularity: \textit{What if, at a given time \( t \), a specific set of popular news articles compete equally for user attention?} As we will explore in subsequent sections, we argue that accurately assessing the popularity of a news article necessitates considering both the \textbf{competition} among news articles for the user's attention and the \textbf{timing} of such competition. Therefore, our method combines \textit{counterfactual reasoning} with a \textit{temporal} approach to adjust the negative sample space, akin to techniques detailed in LANCER \cite{Bae_Ahn_Lee_Kim_2023}.

\section{Method}

\begin{figure}[h]
  \centering
  \includegraphics[width=\linewidth]{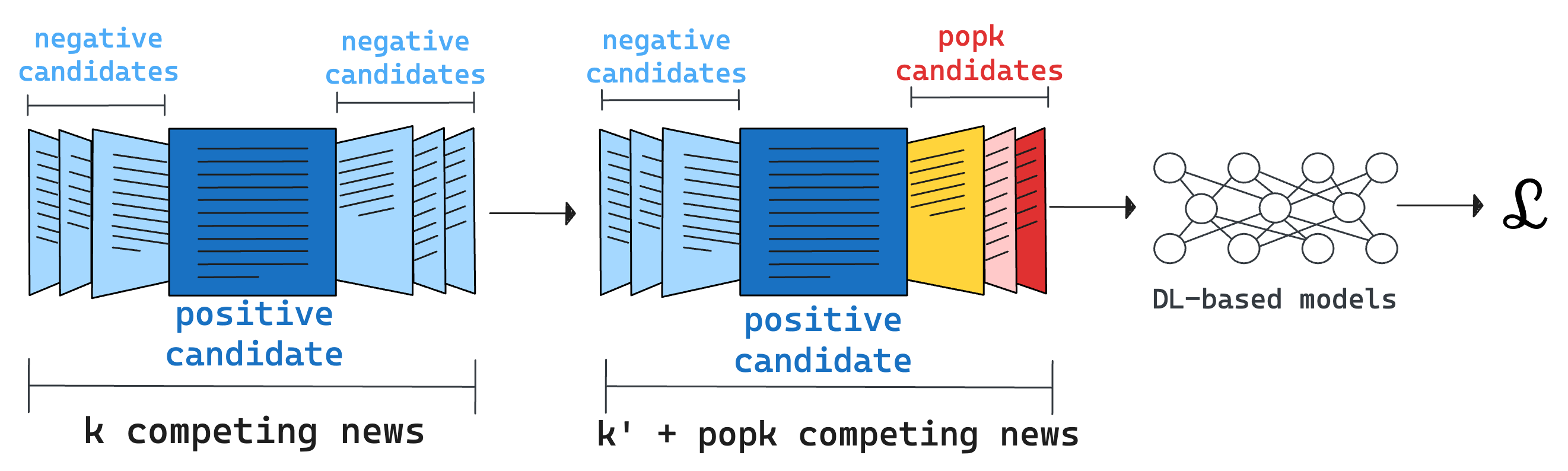}
  \caption{\textit{POPK} overall idea.}
  \label{fig:popk_idea}
\end{figure}

Our proposed method, \textit{POPK}, is based on the idea that popular news articles \textit{\textbf{always}} compete for attention, even if they are not \textit{explicitly} present in the user's impression list. Typically, recommender systems are trained using a negative sampling strategy, where for each positive news article, $k$ negative news articles are selected from the user's impression list. With this approach, the recommender model might become biased towards recommending popular news articles, as these articles are more likely to appear in the candidates sample space due to their higher click rates compared to less popular items. Therefore, the idea behind \textit{POPK} is to counterfactually intervene in the selection of $k$ negative items, as illustrated in Figure \ref{fig:popk_idea}. 

Formally, for each positive candidate news article, we select $k'$ negative news articles where $k' < k$. Additionally, based on a function $\mathcal{P}(t, popk)$, we select the $popk$ most popular news articles within a given time $t$, composing a list of $popk$ news articles, where $popk < k$. Given that \( \textit{popk} = k - k' \), the negative candidate list \( n_{\textit{neg}} \) is adjusted as follows:

\begin{equation}
    n_{\textit{neg}} = k' + \textit{popk} \quad
\end{equation}

\noindent instead of \( n_{\textit{neg}} = k \). The selection of which news articles to replace in the negative candidate list with \( \textit{popk} \) news articles is made \textit{randomly}.

\subsection{Problem Formulation}

Given a user \( u \) and a candidate news article \( n_c \), our objective is to compute an interest score \( Int_{s} \) that quantifies user \( u \)'s engagement potential with the content of \( n_c \). Subsequently, a collection of candidate news articles \( \textbf{N}_c 
 = n_{\textit{pos}} \cup n_{\textit{neg}}\), where $n_{\textit{neg}} = k' + popk$, is evaluated based on the computed interest scores, with the highest-ranking articles recommended to user \( u \). User \( u \) has a historical record of clicked news articles denoted by \( \textbf{H}_u = [n_{u1}, n_{u2}, \ldots, n_{uM}] \), where \( M \) represents the total number of articles previously clicked by the user. Each news article \( n \) in this sequence is characterized by its title \( \mathcal{T} \), abstract \(\mathcal{A}\) and associated entities involved in the news piece \( \textbf{E}_i = [e_1, e_2, \ldots, e_k] \).

\begin{figure}[h]
  \centering
  \includegraphics[width=\linewidth]{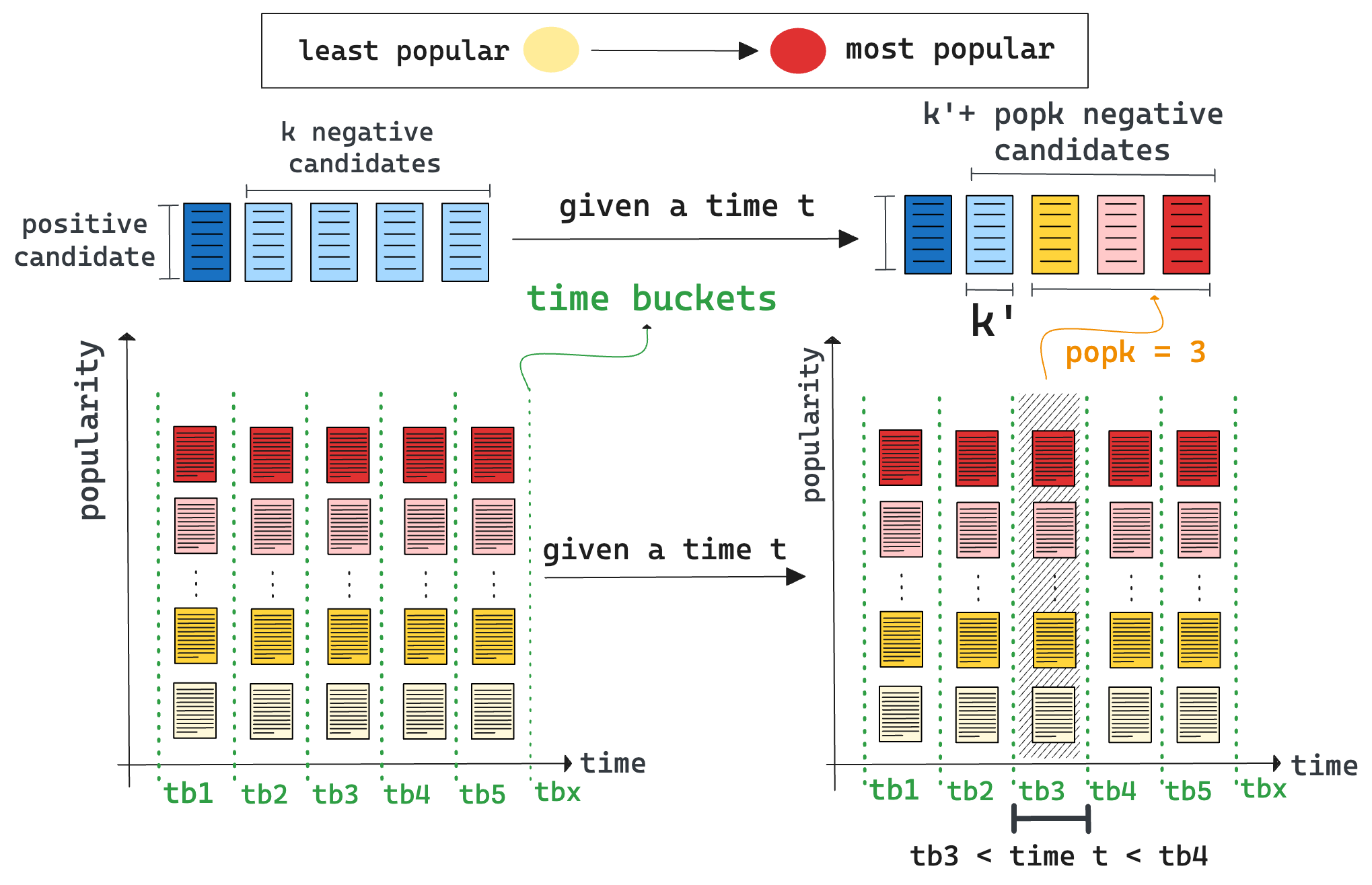}
  \caption{Process of selecting the $popk$ most popular news articles.}
  \label{fig:popk_explain}
  \vspace{-5pt} 
\end{figure}

\subsection{Popular News Articles Selection}

In a simplistic view, a "popular" or "trending" news article is often considered as the one with the highest number of clicks. However, defining a news article as popular involves more nuances, as it encompasses two key aspects: (1) \textit{competition} and (2) \textit{time}. To truly determine the popularity of a news article, we must consider against \textbf{which} news articles it is competing and \textbf{when} it is being evaluated.

\begin{figure}[h]
  \centering
  \includegraphics[width=\linewidth]{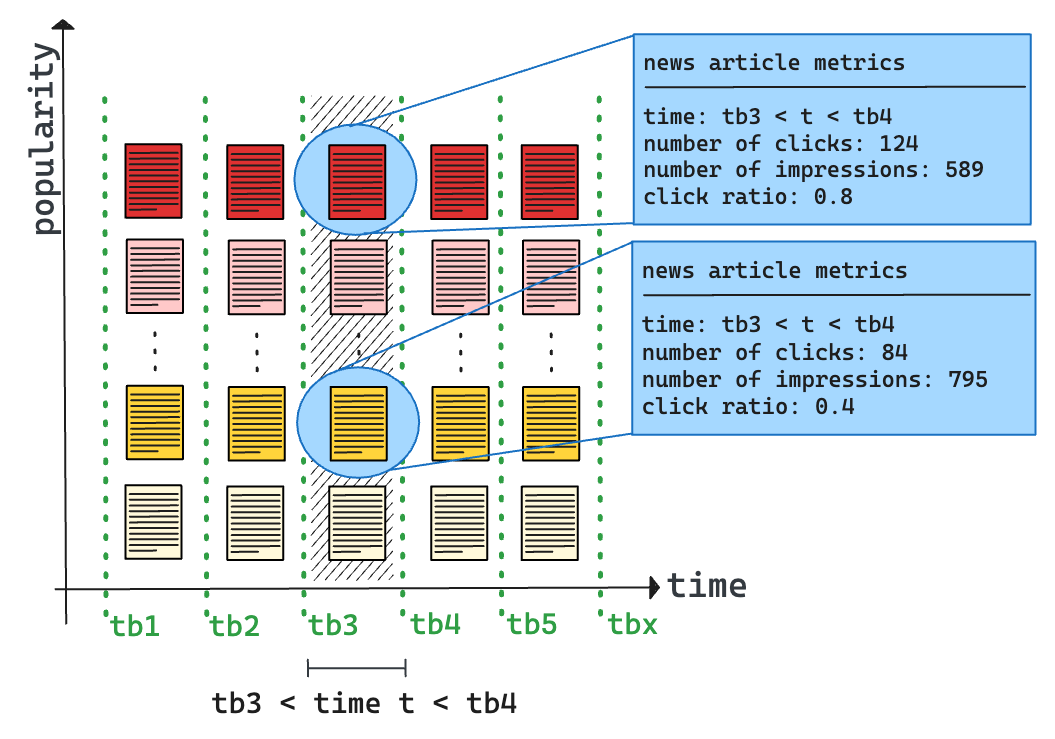}
  \caption{Detailed information per news article at a given time $t$.}
  \label{fig:detail_article_info}
  \vspace{-5pt} 
\end{figure}

Since news articles metrics such as the number of views and number clicks fluctuate over time, we have segmented this trend into what we term as "time buckets." Each time bucket represents a one-hour interval tracking such metrics for each news article. As illustrated in Figure \ref{fig:popk_explain}, each time bucket contains several news articles, each with its corresponding metric values. With the function $\mathcal{P}(t, popk)$, which retrieves the $popk$ most popular news articles at time $t$, we have the flexibility to substitute any given amount of negative candidates with the $popk$ news articles randomly. However, the behavior of the function $\mathcal{P}(t, popk)$ can be understood from different angles. That is, as shown in Figure \ref{fig:detail_article_info}, given a time $t$, we have a broad range of information about all the news articles.

With this information at hand, we can define popularity from different perspectives, and we will consider three for this work. A popular news article is defined as one that:

\begin{enumerate}
    \item $\mathcal{P}(t, popk)$: Has the greatest number of clicks at time $t$.
    \item $\mathcal{P}(t, popk)_{cr}$: Has the highest click ratio at time $t$.
    \item $\mathcal{P}(t, popk)_{var}$: Has the largest variation in clicks within time $t$.
\end{enumerate}

The computation of each of these popularity metrics can follow two distinct approaches: one based on accumulated data (\textit{acc}) and the other on a time-window approach (\textit{ptb}). We will explore both in the following sections.

\subsubsection{\textit{acc} - Accumulated Logic}

The accumulated logic (\textit{acc}) represents the total number of clicks a news article has received up to the analyzed time $t$. The function $\mathcal{P}(t, popk)_{acc}$ retrieves the set of size $popk$ of news articles with the highest accumulated clicks by time $t$. Formally, it selects a set of $popk$ news articles, denoted as $\{a_1, a_2, \ldots, a_{popk}\}$ where each article in the set has received at least as many clicks as the $(popk+1)$-th article at time $t$. In other words, $\mathcal{P}(t, popk)_{acc}$ comprises the top $popk$ articles with the highest accumulated clicks up to time $t$. This is expressed as:

\begin{multline}\label{eq:acc}
    \mathcal{P}(t, popk)_{acc} = \{a_1, a_2, \ldots, a_{popk}\} \mid \\
    \mathcal{C}(a_i, t) \geq \mathcal{C}(a_{popk+1}, t), \forall i \leq popk 
\end{multline}

\noindent
here, $\mathcal{C}(a_i, t)$ retrieves the total number of accumulated clicks that news article $a_i$ has received by time $t$. The same principle applies to $\mathcal{P}(t, popk)_{cr}$ and $\mathcal{P}(t, popk)_{var}$, where $\mathcal{C}(a_i, t)$ would represent the total accumulated click ratio and the accumulated variation in clicks, respectively.

\subsubsection{\textit{ptb} - Per Time Bucket Logic}

The per time bucket logic (\textit{ptb}) represents the total number of clicks a news article has received within a specific time bucket, namely between $t$ and $t - 1 \text{ hour}$. The function $\mathcal{P}(t, popk)_{ptb}$ identifies the set of news articles, of size $popk$, with the highest number of clicks received during this time interval. Formally, it selects a set of $popk$ news articles, denoted as $\{a_1, a_2, \ldots, a_{popk}\}$ where each article in the set has received at least as many clicks as the $(popk+1)$-th article within the time bucket from $t$ to $t-1$ hour. In essence, $\mathcal{P}(t, popk)_{ptb}$ consists of the top $popk$ articles with the highest number of clicks received during the specified time bucket, as shown below:

\begin{multline}\label{eq:ptb}
\mathcal{P}(t, popk)_{ptb} = \{a_1, a_2, \ldots, a_{popk}\} \mid \\
\mathcal{C}(a_i, t, t-1\text{ hour}) \geq \mathcal{C}(a_{popk+1}, t, t-1\text{ hour}), \\
\forall i \leq popk 
\end{multline}

\noindent
here, $\mathcal{C}(a_i, t, t-1)$ denotes the total number of clicks received by news article $a_i$ between time $t$ and $t-1$. The same principle applies to $\mathcal{P}(t, popk)_{cr}$ and $\mathcal{P}(t, popk)_{var}$, where $\mathcal{C}(a_i, t, t-1)$ would represent the total click ratio between time $t$ and $t-1$ and the variation in clicks between time $t$ and $t-1$, respectively.

\subsubsection{Example}

Let's illustrate an example of how \textit{POPK} works. Consider a user $u$ with an impression list of news articles $\{n_{1245}-0, n_{12}-0, n_{125}-0, n_{45}-0, n_{90}-0, n_{1289}-0, n_{546}-0, n_{174}-1\}$, where "$-0$" indicates the user did not click the article and "$-1$" indicates a click. Traditionally, as implemented in the NAML model \cite{wu2019neural}, a constant number $k$ of non-clicked articles is chosen alongside the positive article. For instance, choosing $k = 4$, the training set would be $n_{pos} = \{n_{174}-1\}$ and choosing \textit{randomly} the negative news articles $n_{neg} = \{n_{125}-0, n_{90}-0, n_{546}-0, n_{12}-0\}$. 

Here's where \textit{POPK} comes into play. We add an additional step to this process of generating training samples. After randomly selecting $k = 4$ negative articles, we select a constant number of popular articles, for example $popk = 2$, to include in this set of negative samples. Suppose, for example, that the impression log is from $t = 17$th January 2017 ($t_{17}$). Using either the $acc$ or $ptb$ logic, $\mathcal{P}(t_{17}, popk = 2)$ might yield $\{n_{19}-0, n_{70}-0\}$, where $n_{19}$ and $n_{70}$ are the 2 most popular news articles based on the \textit{number of clicks} at this given time $t_{17}$. Consequently, $k' = k - popk = 4 - 2 = 2$. We then randomly select $k'$ articles from $n_{neg}$. Let's say $n_{neg}^{k'} = \{n_{125}-0, n_{90}-0\}$ and add these two popular articles, resulting in $n_{neg} = \{n_{125}-0, n_{90}-0, n_{19}-0, n_{70}-0\}$. Finally, we arrive at the candidate list of news articles \( n_{pos} = \{ n_{174}-1 \} \cup n_{neg} = \{ n_{125}-0, n_{90}-0, n_{19}-0, n_{70}-0 \} \), where \( n_{neg} \) now includes the 2 most popular articles at the given time \( t_{17} \). 

Through this process, applying the aforementioned popularity measurement described by equations \ref{eq:acc} and \ref{eq:ptb}, we can systematically eliminate the \textbf{implicit} influence of popular news articles during each training step. By doing so, our recommender model can more \textbf{explicitly} understand that the clicked news article \( n_{174} \) was indeed competing with popular articles such as \( n_{19} \) and \( n_{70} \). This approach allows the model to better discern that the user prefers topics like \( n_{174} \) over popular news articles.

\subsection{Loss Function}

We use a metric-learning objective, minimizing cross-entropy loss and incorporating negative sampling, inspired by the NRMS model \cite{wu-etal-2019-neural-news}. For each clicked news item (positive sample), we randomly sample \( n_{\textit{neg}} = k' + popk \) non-clicked news items (negative samples) from the same impression. Let \( \hat{y}^+ \) be the click probability score of the positive news item, and \([\hat{y}^-_1, \hat{y}^-_2, \ldots, \hat{y}^-_K]\) be the scores for the \( n_{\textit{neg}} \) negative items. These scores are normalized using softmax to compute the posterior click probability of a positive sample: 

\begin{equation}
    p_i = \frac{\exp(\hat{y}^+_i)}{\exp(\hat{y}^+_i) + \sum_{j=1}^{K} \exp(\hat{y}^-_{i,j})}
\end{equation}

We reformulate the task as a pseudo \((n_{\textit{pos}} = n_{\textit{neg}} + 1)\)-way classification problem. The loss function for training is the negative log-likelihood of all positive samples \( S \): $\mathcal{L} = - \sum_{i \in S} \log(p_i)$. This approach effectively distinguishes between clicked and non-clicked news items within the same impression.

\section{Experiment}

\subsection{Baseline Models}

All experiments were conducted using the framework \textbf{\textit{newsreclib}} \cite{iana2023newsreclib}. We integrated \textit{POPK} into the following baseline models: (1) \textbf{NRMS} \cite{Wu_2019}: Utilizes a multi-head self-attention network to learn user and news representations; (2) \textbf{NAML} \cite{wu2019neural}: Employs a multi-view attention network to learn representations based on various news article features. (3) \textbf{LSTUR} \cite{an-etal-2019-neural}: Jointly models users' long-term and short-term interests using a GRU network.

\subsection{Datasets} We assessed the performance of \textit{POPK} using three datasets: \textit{MIND-small} (in English) \cite{wu-etal-2020-mind}, \textit{Adressa one-week} (in Norwegian) \cite{10.1145/3106426.3109436}, and a proprietary dataset, referred to as \textit{Nikkei}, which is in Japanese. The proprietary dataset is from the company Nikkei Inc.\footnote{\url{https://www.nikkei.com/}}, based in Japan, which is a leading media corporation renowned for its comprehensive coverage of business, economic, and financial news. The statistics for each training, validation, and testing split used in our work are summarized in Table \ref{tab:dataset_stats}.

For the \textit{Nikkei} dataset, we have a total of 15,803 news articles. The user click behavior data was divided into training (January 16, 2023, to January 20, 2023), validation (January 21, 2023), and testing (January 22, 2023) sets. This dataset was constructed by randomly sampling paid users, excluding noises such as unacceptable crawlers and users with unnatural large browser counts. The \textit{Adressa one-week} dataset contains 22,136 news articles. We partitioned the user click behavior data into training (January 1, 2017, to January 5, 2017), validation (January 6, 2017), and testing (January 7, 2017) sets. For the \textit{MIND-small} dataset, there are 93,698 news articles. The user click behavior data was divided into training (November 9, 2019, to November 13, 2019), validation (November 14, 2019), and testing (November 15, 2019) sets.

\begin{table}[htbp]
\centering
\small
\caption{Dataset Statistics}
\label{tab:dataset_stats}
\begin{tabular}{@{}cccc@{}}
\toprule
\textbf{Dataset} & \textbf{Train} & \textbf{Validation} & \textbf{Test} \\ \midrule
 & Impr./Users & Impr./Users & Impr./Users \\ \midrule
\textit{Adressa} & 181,279 / 83,599 & 36,412 / 27,943 & 145,626 / 68,565 \\
\textit{MIND-small} & 124,229 / 45,214 & 29,498 / 19,703 & 70,938 / 48,593 \\
\textit{Nikkei} & 137,142 / 23,139 & 10,560 / 6,201 & 9,695 / 5,805 \\ \bottomrule
\end{tabular}
\end{table}

\begin{table*}[htbp]
\centering
\footnotesize 
\resizebox{\textwidth}{!}{ 
\begin{tabular}{|l|l|cccc|cccc|cccc|}
\hline
\multirow{4}{*}{\textit{\textbf{Nikkei}}} & \multirow{4}{*}{} & \multicolumn{4}{c|}{NRMS} & \multicolumn{4}{c|}{NAML} & \multicolumn{4}{c|}{LSTUR} \\ \cline{3-14} 
                  & metrics & AUC    & MRR   & $nDCG@5$ & $nDCG@10$ & AUC    & MRR   & $nDCG@5$ & $nDCG@10$ & AUC    & MRR   & $nDCG@5$ & $nDCG@10$ \\ \cline{2-14}
{} & \textit{Original}  & 0.5140 & \underline{0.3110} & \underline{0.2117} & 0.3069 & 0.5011 & 0.3323 & 0.2324 & 0.3322 & \textbf{0.4953} & 0.2878 & 0.1929 & 0.2883   \\ \hline
\multirow{3}{*}{$acc$} & $popk$ = 1 & 0.4993 & 0.2999 & \textbf{0.2137} & 0.3072  & \underline{0.5411} & \textbf{0.3723} & \textbf{0.2790} & \textbf{0.3780}  & 0.4907 & \underline{0.3097} & 0.2156 & 0.3065   \\ 
                     & $popk$ = 2 & 0.4854 & 0.3027 & 0.2115 & \textbf{0.3217} & 0.5031 & 0.3408 & 0.2416 & 0.3419 & 0.4936 & \textbf{0.3190} & \textbf{0.2303} & 0.3289   \\ 
                     & $popk$ = 3 & 0.5000 & 0.3000 & 0.1986 & 0.2977 & 0.5002 & 0.3038 & 0.2160 & 0.2996 & \textbf{0.4939} & 0.3085 & \underline{0.2189} & \textbf{0.3135}  \\ \hline
\multirow{3}{*}{$ptb$} & $popk$ = 1 & \underline{0.5225} & 0.2892 & 0.1854 & 0.2708 & 0.5258 & 0.3346 & 0.2419 & 0.3415 & 0.4913 & 0.2875 & 0.1939 & 0.2908  \\ 
                     & $popk$ = 2 & \textbf{0.5234} & \textbf{0.3133} & 0.2058 & \underline{0.3108} & \textbf{0.5500} & \underline{0.3516} & \underline{0.2570} & \underline{0.3511} & 0.4902 & 0.2845 & 0.1870 & 0.2761   \\ 
                     & $popk$ = 3 & 0.5000 & 0.2917 & 0.1993 & 0.2942 & 0.5316 & 0.3274 & 0.2360 & 0.3355 & 0.4915 & 0.3086 & 0.2176 & \underline{0.3123}   \\ \hline
\rowcolor{lightgray}
\multicolumn{2}{|c|}{Increase vs \textit{Original}} & 1.82\% & 0.73\% & 0.94\% & 4.82\% & 9.75\% & 12.03\% & 20.05\% & 13.78\% & -0.28\% & 7.11\% & 19.38\% & 8.74\% \\ \hline

\multirow{4}{*}{\textit{\textbf{MIND-small}}} & \multirow{4}{*}{} & \multicolumn{4}{c|}{NRMS} & \multicolumn{4}{c|}{NAML} & \multicolumn{4}{c|}{LSTUR} \\ \cline{3-14} 
                  & metrics & AUC    & MRR   & $nDCG@5$ & $nDCG@10$ & AUC    & MRR   & $nDCG@5$ & $nDCG@10$ & AUC    & MRR   & $nDCG@5$ & $nDCG@10$ \\ \cline{2-14}
{} & \textit{Original} & 0.5000 & 0.2731 & 0.2532 & 0.3198 & 0.5000 & 0.2806 & 0.2606 & 0.3246 & 0.5000 & 0.3000 & 0.2830 & 0.3468   \\ \hline
\multirow{3}{*}{$acc$} & $popk$ = 1 & 0.5000 & \textbf{0.3009} &\textbf{ 0.2858} & \textbf{0.3498} & 0.5000 & 0.2429 & 0.2192 & 0.2819  & 0.5000 & \underline{0.3025} & 0.2847 & 0.3480   \\ 
                     & $popk$ = 2 & 0.5000 & 0.2711 & 0.2499 & 0.3136 & 0.5007 & \textbf{0.3177} & \textbf{0.2997} & \textbf{0.3623}   & 0.5000 & \textbf{0.3038} & \textbf{0.2862} & \textbf{0.3503}   \\ 
                     & $popk$ = 3 & 0.5000 & \underline{0.2866} & \underline{0.2709} & \underline{0.3354} & \textbf{0.5417} & \underline{0.3119} & \underline{0.2950} & \underline{0.3578}  & 0.5000 & 0.3006 & \underline{0.2851} & \underline{0.3487}   \\ \hline
\multirow{3}{*}{$ptb$} & $popk$ = 1 & 0.5000 & 0.2467 & 0.2230 & 0.2858 & \underline{0.5034} & 0.2816 & 0.2659 & 0.3322  & 0.5000 & 0.2898 & 0.2732 & 0.3364   \\ 
                     & $popk$ = 2 & 0.5000 & 0.2131 & 0.1894 & 0.2535 & 0.5000 & 0.2479 & 0.2265 & 0.2908 & 0.5000 & 0.2883 & 0.2693 & 0.3345   \\ 
                     & $popk$ = 3 & 0.5000 & 0.2350 & 0.2128 & 0.2760 & 0.5000 & 0.2449 & 0.2215 & 0.2846 & 0.5000 & 0.2962 & 0.2800 & 0.3423   \\ \hline
\rowcolor{lightgray}
\multicolumn{2}{|c|}{Increase vs \textit{Original}} & 0.00\% & 10.12\% & 12.87\% & 9.38\% & 8.33\% & 13.22\% & 15.00\% & 11.61\% & 0.00\% & 1.26\% & 1.13\% & 1.00\% \\ \hline

\multirow{4}{*}{\textit{\textbf{Adressa one-week}}} & \multirow{4}{*}{} & \multicolumn{4}{c|}{NRMS} & \multicolumn{4}{c|}{NAML} & \multicolumn{4}{c|}{LSTUR} \\ \cline{3-14} 
                  & metrics & AUC    & MRR   & $nDCG@5$ & $nDCG@10$ & AUC    & MRR   & $nDCG@5$ & $nDCG@10$ & AUC    & MRR   & $nDCG@5$ & $nDCG@10$ \\ \cline{2-14}
{} & \textit{Original} & \textbf{0.5847} & 0.2102 & 0.1824 & 0.2894  & 0.5000 & 0.2267 & 0.1984 & 0.2945  & 0.5129 & 0.3099 & 0.2666 & 0.3350   \\ \hline
\multirow{3}{*}{$acc$} & $popk$ = 1 & 0.5039 & 0.2122 & 0.1363 & 0.2078  & \underline{0.5767 }& 0.2834 & 0.2830 & 0.3647 & \underline{0.5422} & \underline{0.3097} & 0.2836 & 0.3517   \\ 
                     & $popk$ = 2 & 0.5000 & 0.2265 & 0.1392 & 0.2131 & 0.5002 & \textbf{0.3512} & \textbf{0.3370} & \textbf{0.4066} & 0.5191 & \underline{0.3411} & \underline{0.3158} & \underline{0.3776}  \\ 
                     & $popk$ = 3 & 0.5000 & 0.2485 & 0.1733 & 0.2441 & \textbf{0.5805} & \underline{0.3267} & \underline{0.3201} & \underline{0.3688} & \textbf{0.5705} & \textbf{0.3551} & \textbf{0.3283} & \textbf{0.3896} \\ \hline
\multirow{3}{*}{$ptb$} & $popk$ = 1 & \underline{0.5402} & \textbf{0.2766} & \underline{0.2720} & \textbf{0.3500}  & 0.5575 & 0.2546 & 0.2657 & 0.3336 & 0.5270 & 0.2416 & 0.2389 & 0.3126   \\ 
                     & $popk$ = 2 & 0.5145 & \underline{0.2745} & \textbf{0.2735} & \underline{0.3389} & 0.5529 & 0.2429 & 0.2493 & 0.3156 & 0.5315 & 0.2249 & 0.2193 & 0.2858   \\ 
                     & $popk$ = 3 & 0.5098 & 0.2561 & 0.2568 & 0.3233 & 0.5433 & 0.2428 & 0.2426 & 0.3050 & 0.5235 & 0.2077 & 0.2055 & 0.2789  \\ \hline
\rowcolor{lightgray}
\multicolumn{2}{|c|}{Increase vs \textit{Original}} & -7.61\% & 31.59\% & 49.95\% & 20.94\% & 16.10\% & 54.92\% & 69.86\% & 38.06\% & 11.23\% & 14.59\% & 23.14\% & 16.30\% \\ \hline
\end{tabular}
}
\caption{Performance Metrics for \textit{Nikkei}, \textit{MIND-small}, and \textit{Adressa one-week} Datasets. We've identified in \textbf{bold} the best-performing metrics and indicated with \underline{underline} the second-best. In cases where the differences were marginal, emphasis were added to similar values.}
\label{tab:results}
\end{table*}

\subsection{Metrics}

In line with prior research \cite{qi2021pprec, Yang_2023}, we evaluate model performance using Area under the ROC Curve (AUC), Mean Reciprocal Rank (MRR), and normalized discounted cumulative gain ($nDCG@5$ and $nDCG@10$). In addition, inspired by \cite{iana2024train}, we decided to employ additional metrics that help understand how diverse our recommendations are. That is, we assess aspect-based category diversity at position \(k\) using normalized entropy:

\begin{equation}
    D_{ctg}@k = -\frac{\sum_{j \in D_{ctg}} p(j)\log p(j)}{\log(|D_{ctg}|)}    
\end{equation}

\noindent
where \(D_{ctg}\) denotes the number of different categories in each dataset.

\subsection{\textit{POPK} Configuration}

For each baseline, we conducted experiments with a fixed number of negative candidates $n_{\textit{neg}} = 4$ for every positive candidate. We varied the parameters $k'$ from 3 to 1 and $popk$ from 1 to 3, always maintaining $k$ at a fix value of $k = n_{\textit{neg}} = 4$.

\subsection{Text Embeddings and Model Training}

In our process to evaluate \textit{POPK} we followed the methodologies outlined in \cite{iana2023newsreclib, iana2024train}. Specifically, we utilized RoBERTa Base \cite{liu2019roberta}, a monolingual English Pre-trained Language Model (PLM) for the \textit{MIND-small}, NB-BERT Base \cite{kummervold-etal-2021-operationalizing}, a monolingual Norwegian PLM for \textit{Adressa one-week}, and a finetuned version of the DeBERTa model \citep{he2021deberta} called \textit{Japanese DeBERTa V2}\footnote{\url{https://huggingface.co/ku-nlp/deberta-v2-base-japanese}} for the \textit{Nikkei} dataset. For all the models, we adopted a training strategy that involves fine-tuning only the last four layers of the PLM. We set the maximum history length to 50 and tuned all other model-specific hyperparameters for the baselines to optimal values as reported in the relevant literature. During training, we employed mixed precision and optimized using the Adam algorithm \cite{kingma2017adam}. The learning rate was set to 1e-5 for the \textit{MIND-small} and \textit{Nikkei} datasets and 1e-6 for the \textit{Adressa one-week} dataset and we trained all models for 3 epochs using a single NVIDIA A100 GPU. 


\section{Results}

In this section, we comprehensively evaluate \textit{POPK} by answering the following research questions:

\begin{itemize}
    \item \textbf{RQ1} (Accuracy): How much does \textit{POPK} enhance current methods in personalized news recommendation?
    \item \textbf{RQ2} (Diversity): To what extent does \textit{POPK} improve current methods in terms of diversity?
    \item \textbf{RQ3} (Trade-off Between Accuracy and Diversity): Is there a balance of \textit{popk} values between accuracy and diversity?
    \item \textbf{RQ4} (Influence of Popularity Measurement): How does altering the popularity measurement metric affect model performance in terms of accuracy?
    \item \textbf{RQ5} (Limitations and Advantages): What are the pros and cons of the proposed \textit{POPK} method?
\end{itemize}

\subsection{RQ1. Accuracy}

\subsubsection{Overview}

Our findings across the three datasets and the three baseline models are summarized in Table \ref{tab:results}. We evaluate both the original and modified versions of the models, varying $popk$ from 1 to 3 according to two logics: $acc$ and $ptb$. Additionally, the table includes a row labeled \textit{"Increase vs Original,"} showcasing the improvement compared to the original methods. Results are based on $\mathcal{P}(t, popk)$, identifying popular news articles based on the \textit{number of clicks}.

\subsubsection{\textit{Nikkei}}

Consistent and significant improvements were observed across all baseline models. In this dataset, the \textit{acc} strategy consistently outperformed \textit{ptb}, with the exception of NRMS, which exhibited improvements in AUC and MRR metrics with \textit{ptb}. While AUC underperformed compared to the original model in LSTUR, it provided enhanced MRR, nDCG@5, and nDCG@10 scores by up to about 7.11\%, 19.38\%, and 8.74\%, respectively.

\subsubsection{\textit{MIND-small}}

Consistent improvements were observed across different models for \textit{MIND-small}. Though LSTUR exhibited the least increment, this shortfall was compensated by a significant increase in recommendation diversity, as discussed later. Specifically, accuracy enhancements predominantly resulted from the \textit{acc} logic rather than \textit{ptb}. For instance, the NAML model showed improvements in AUC by up to 8.33\%, MRR by 13.22\%, nDCG@5 by 15\%, and nDCG@10 by 11.61\%.

\subsubsection{\textit{Adressa one-week}}

The model yielded substantial enhancements for this dataset, with the exception of the AUC metric with the NRMS baseline. Notably, the \textit{acc} logic demonstrated superior performance for NAML and LSTUR, while the \textit{ptb} logic proved more effective for the NRMS which provided enhanced MRR, nDCG@5, and nDCG@10 scores by up to about 31.59\%, 49.95\%, and 20.94\%, respectively.


\subsubsection{General Observations}

By applying our \textit{POPK} method, we augment baseline models' ability to accurately discern user-preferred news articles. For instance, during the World Cup, a user accessing a news platform is likely aware of the championship due to extensive exposure from diverse media platforms but may never click on such news articles. If the negative samples do not include articles about this topic, the model might struggle to understand that the user is not seeking a deeper understanding of it. 

\begin{figure}[h]
  \centering
  \includegraphics[width=\linewidth]{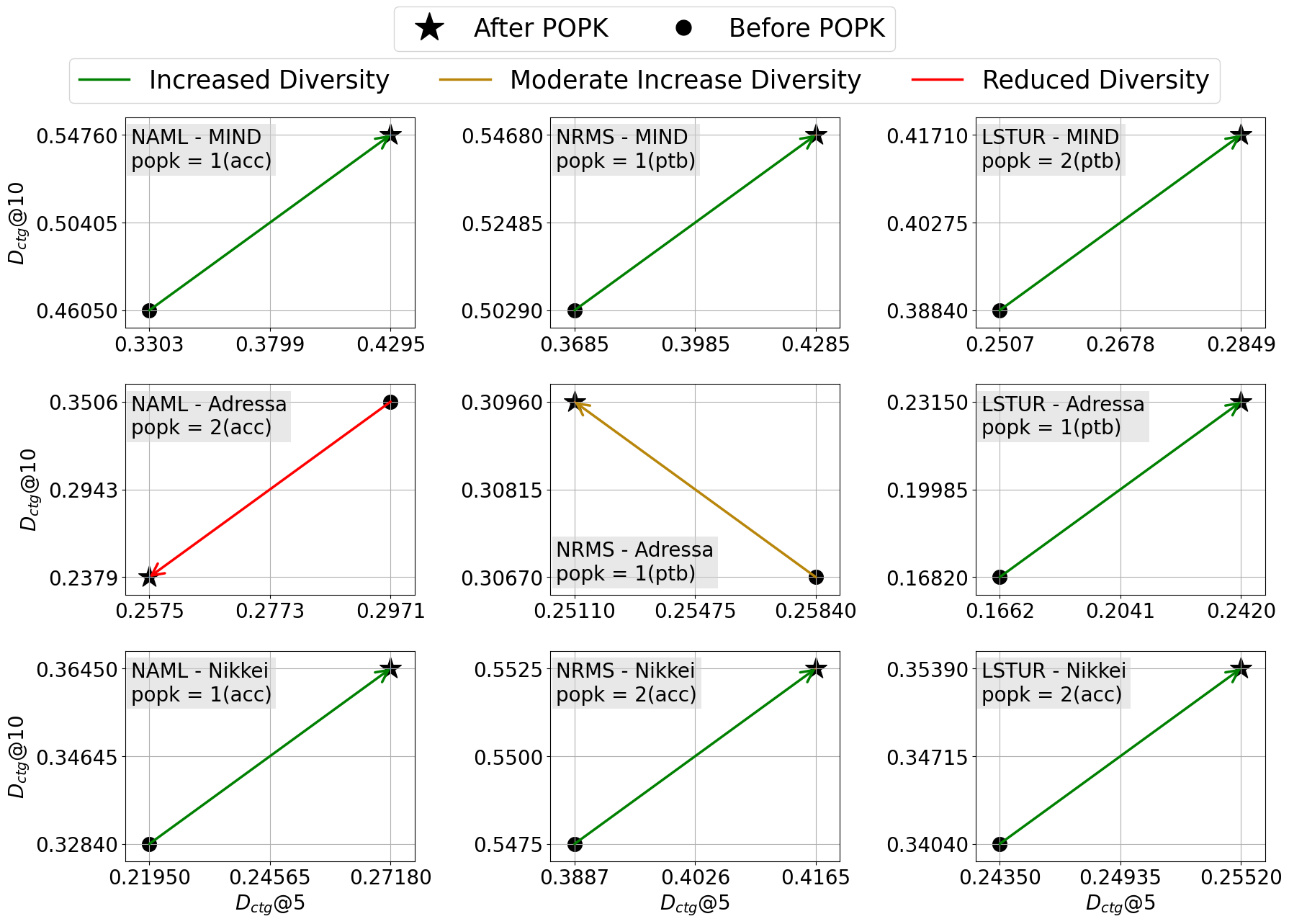}
  \caption{Diversity increase $D_{ctg}@5/10$ between original baseline model and \textit{POPK} version.}
  \label{fig:diversity_increase}
  \vspace{-10pt} 
\end{figure}

We affirm that popular news articles are \textit{always} \textbf{implicitly} competing for the user's attention. \textit{POPK} mitigates this implicit influence thereby refining user interest comprehension and enhancing model accuracy. Furthermore, variations in $popk$ contribute to either increased diversity or accuracy, as discussed further in Sections \ref{sec:diversity} and \ref{sec:tradeoff}. 

In the case of the \textit{Adressa one-week} dataset, which includes information on click patterns for news articles published long ago (e.g., a user clicking on a news article from the 2000s during the early weeks of 2017), consider a scenario where a user logs into the news platform during the launch of a new iPhone in 2017, causing tech news articles to experience a surge in popularity. Surprisingly, the user clicks on a 2006 news article about politics instead. Thus, what \textit{POPK} does is reduce the \textbf{implicit} impact of the iPhone news articles and add it into the candidate news articles \textbf{explicitly}. Given that such articles are highly popular, we argue that the user is aware of their existence but chooses not to engage with them, indicating a preference for politics over tech news. While this example is simplified, it effectively illustrates how incorporating a specified number of popular news articles, \( popk \), into the list of negative candidate news articles helps the model better understand user preferences.

\subsection{RQ2. Diversity}\label{sec:diversity}

\subsubsection{Diversity Metrics}

To explore recommendation diversity further, we conducted a comprehensive analysis comparing baseline models with their respective modified versions. These modifications incorporated both the \( \text{acc} \) and \( \text{ptb} \) logics, while varying \textit{popk} from 1 to 3. The best-performing model, on average, was selected for this comparative assessment, illustrated in Figure \ref{fig:diversity_increase} by the \( D_{ctg}@5 \) vs \( D_{ctg}@10 \) graph. All results were computed using \( \mathcal{P}(t, popk) \), which identifies popular news articles based on the \textit{number of clicks}. The figure displays results before and after applying \textit{POPK} to each baseline model using black circle and and black star, respectively. Points are connected with arrows: green indicates increased diversity if both \( D_{ctg}@5 \) and \( D_{ctg}@10 \) increased, yellow if only \( D_{ctg}@10 \) increased, and red if both decreased. The figure illustrates a notable enhancement in recommendation category diversity across most models and datasets. However, for \textit{Adressa one-week} dataset the NRMS model for popk = 1(ptb) showed a moderate increase, and NAML model popk = 2(acc) exhibited decreased category diversity. Significant improvements were consistently observed across other methods, particularly on the \textit{Nikkei} and \textit{MIND-small} datasets, where enhancements ranged from 0.91\% to 30\%. Notably, the LSTUR model demonstrated a substantial increase of 45.62\% on the \textit{Adressa one-week} dataset.

\subsubsection{Categories Frequencies}

\begin{figure}[h]
  \centering
  \includegraphics[width=\linewidth]{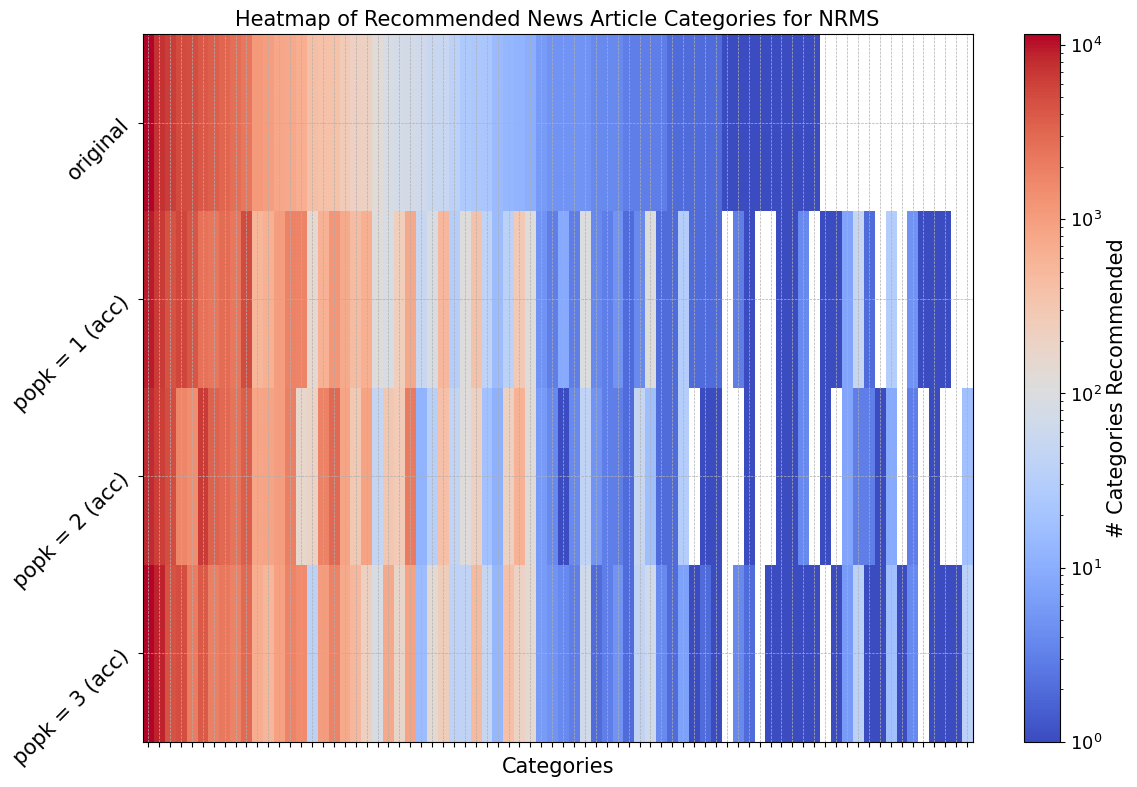}
  \caption{Category log-heatmap NRMS model ($popk$ = 1 to 3, \textit{acc}).}
  \label{fig:heat_map_nrms}
  \vspace{-10pt} 
\end{figure}

\begin{figure}[h]
  \centering
  \includegraphics[width=\linewidth]{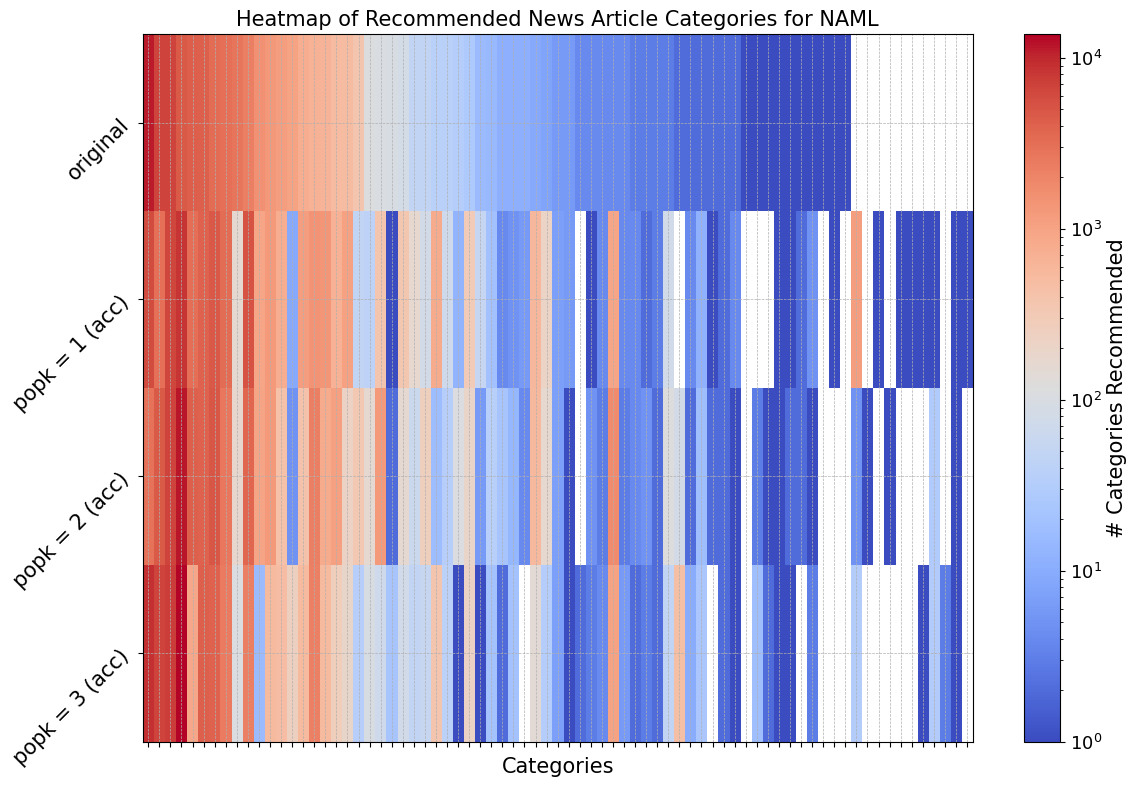}
  \caption{Category log-heatmap NAML model ($popk$ = 1 to 3, \textit{acc}).}
  \label{fig:heat_map_naml}
  \vspace{-10pt} 
\end{figure}

\begin{figure}[h]
  \centering
  \includegraphics[width=\linewidth]{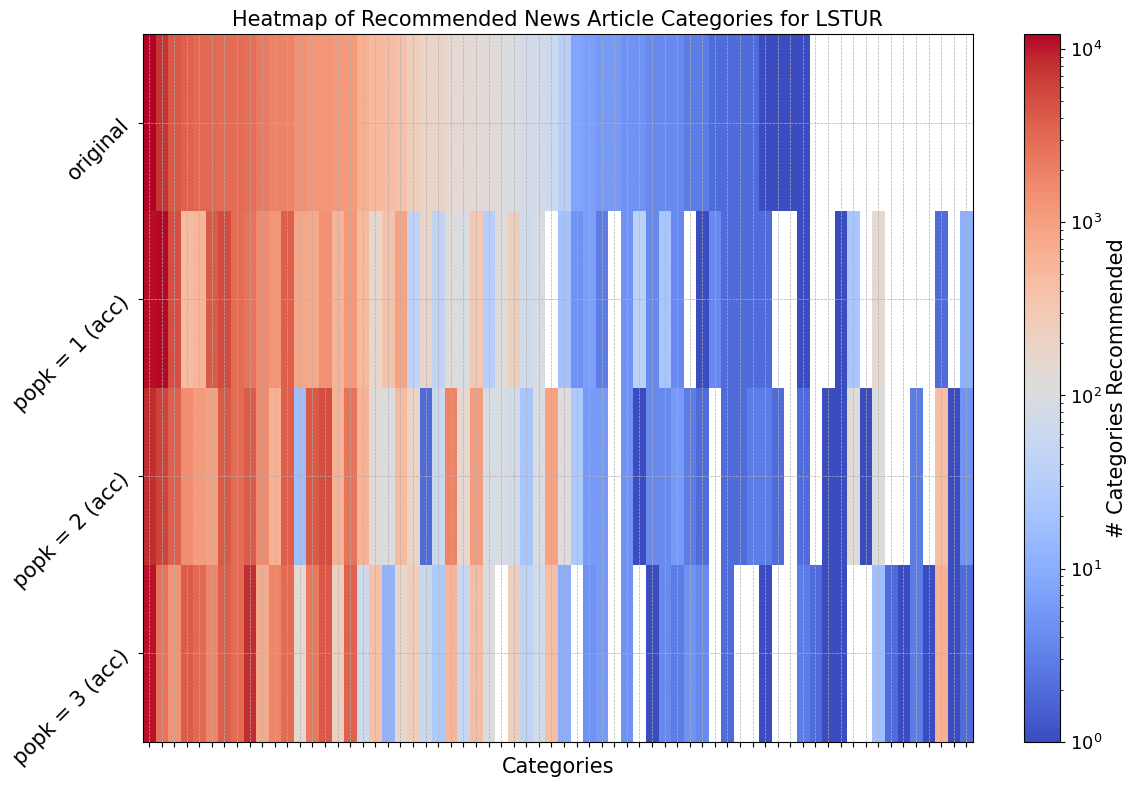}
  \caption{Category log-heatmap LSTUR model ($popk$ = 1 to 3, \textit{acc}).}
  \label{fig:heat_map_lstur}
\end{figure}

To delve even deeper into our analysis, we generated log-heatmaps illustrating the frequency distribution of recommended categories for the \textit{Nikkei} dataset across all baseline models (NRMS, NAML, and LSTUR) using the \textit{POPK} logic with \textit{acc}. Figures \ref{fig:heat_map_nrms}, \ref{fig:heat_map_naml}, and \ref{fig:heat_map_lstur} depict these distributions respectively. The heatmaps categorize news article recommendations based on their frequency across models, where darker red indicates higher frequency, darker blue indicates lower frequency and white indicates zero recommendation. The first column of each heatmap displays the frequency of recommended categories from right (most recommended - dark red) to left (least recommended - dark blue and white) for each baseline model (NRMS, NAML, and LSTUR) prior to applying the \textit{POPK} method. Our analysis demonstrates that \textit{POPK} significantly enhances the recommendation of long-tail items. Compared to the baseline models, our approach recommends a broader range of items. 

Notably, in the \textit{POPK} models, categories on the right side of the graphs are shaded in blue, indicating increased recommendations for articles that were not recommended at all by the baseline (represented in white). Conversely, the extreme left of our \textit{POPK} models shows less intense dark red colors, sometimes even closer to grey, which represents the midpoint of the recommended categories' color scale. This suggests a shift towards recommending a more diverse set of items rather than primarily focusing on a few popular ones, as observed in the original baselines.




\subsection{RQ3. Trade-off Between Accuracy and Diversity}\label{sec:tradeoff}

\begin{figure*}[h]
  \centering
  \includegraphics[width=0.8\linewidth]{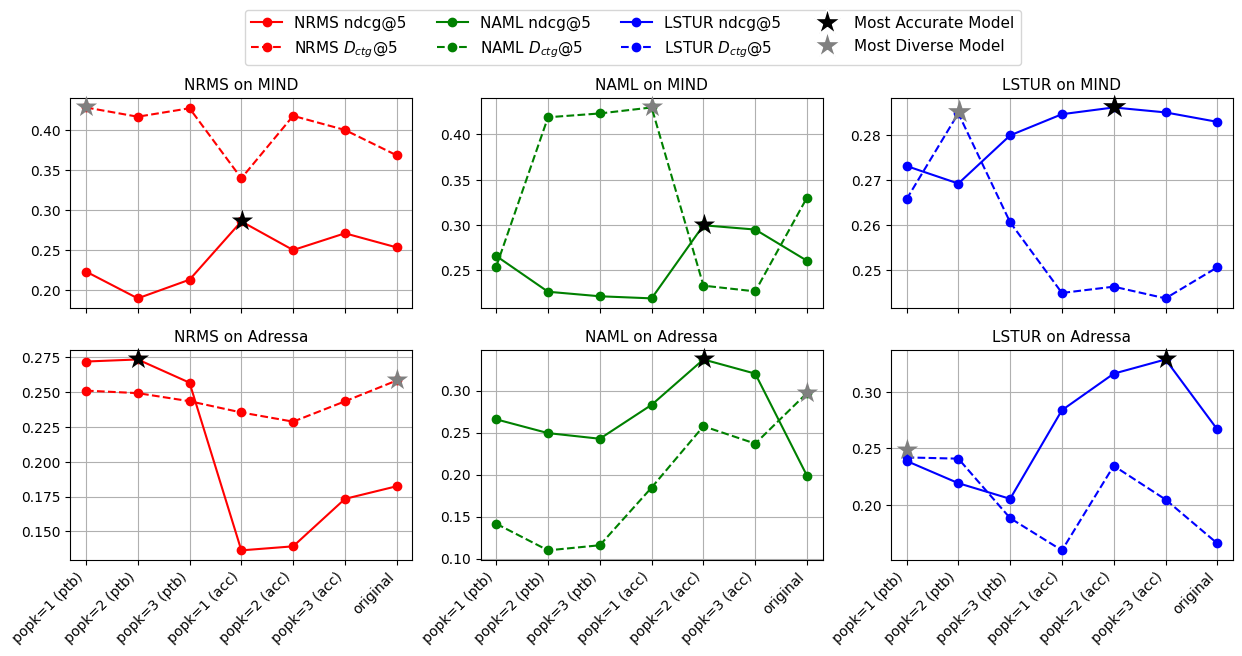}
  \caption{Tradeoff - Accuracy and Category Diversity for the \textit{MIND-small} and \textit{Adressa one-week} datasets}
  \label{fig:tradeoff}
\end{figure*}

It is notable that our analysis reveals intriguing dynamics concerning the trade-off between accuracy and diversity when employing \textit{POPK}. This relationship is depicted in Figure \ref{fig:tradeoff}, which presents findings from experiments conducted on the \textit{MIND-small} and \textit{Adressa one-week} datasets using the baseline models. We evaluated the original models and their modified version under both $ptb$ and $acc$ logics while varying the $popk$ parameter from 1 to 3. The analysis relies on $\mathcal{P}(t, popk)$, which identifies popular news articles based on the \textit{number of clicks}. The graph uses solid lines for \( nDCG@5 \), dashed lines for \( D_{ctg}@5 \), and grey and black stars for the most diverse and accurate models, respectively.

\subsubsection{\textit{MIND-small}}
The alternating pattern of dashed and solid lines illustrates an inverse correlation between the diversity of recommended topics and the accuracy of recommendations. For example, observe the behavior of the NAML model (blue lines): increasing \textit{popk} according to the $ptb$ logic enhances diversity scores but diminishes accuracy. Conversely, following the $acc$ logic, increasing \textit{popk} enhances accuracy but reduces diversity.

\subsubsection{\textit{Adressa one-week}}
Conversely, for the \textit{Adressa}, we observe both inverse and proportional relationships at different points. Consider the behavior of the LSTUR model (blue line): initially under the $ptb$ logic, increasing \textit{popk} leads to decreased accuracy but increased diversity. However, for \textit{popk = 3} (ptb) and all values of \textit{popk} under the $acc$ logic, both dashed and solid lines move in the same direction, indicating concurrent increases or decreases in accuracy and diversity.

\begin{table*}[htbp]
\centering
\Large 
\renewcommand{\arraystretch}{2}
\resizebox{\textwidth}{!}{ 
\begin{tabular}{|l|>{\columncolor[gray]{0.95}}c|>{\columncolor[gray]{0.92}}c|>{\columncolor[gray]{0.89}}c|>{\columncolor[gray]{0.95}}c|>{\columncolor[gray]{0.92}}c|>{\columncolor[gray]{0.89}}c|>{\columncolor[gray]{0.95}}c|>{\columncolor[gray]{0.92}}c|>{\columncolor[gray]{0.89}}c||>{\columncolor[gray]{0.95}}c|>{\columncolor[gray]{0.92}}c|>{\columncolor[gray]{0.89}}c|>{\columncolor[gray]{0.95}}c|>{\columncolor[gray]{0.92}}c|>{\columncolor[gray]{0.89}}c|>{\columncolor[gray]{0.95}}c|>{\columncolor[gray]{0.92}}c|>{\columncolor[gray]{0.89}}c||>{\columncolor[gray]{0.95}}c|>{\columncolor[gray]{0.92}}c|>{\columncolor[gray]{0.89}}c|>{\columncolor[gray]{0.95}}c|>{\columncolor[gray]{0.92}}c|>{\columncolor[gray]{0.89}}c|>{\columncolor[gray]{0.95}}c|>{\columncolor[gray]{0.92}}c|>{\columncolor[gray]{0.89}}c|}
\hline
 \multirow{2}{*}{\textit{Popularity Study}} & \multicolumn{9}{c||}{NRMS} & \multicolumn{9}{c||}{NAML} & \multicolumn{9}{c|}{LSTUR} \\ \cline{2-28}
 & \multicolumn{3}{c|}{$\mathcal{P}$} & \multicolumn{3}{c|}{$\mathcal{P}_{cr}$} & \multicolumn{3}{c||}{$\mathcal{P}_{var}$} & \multicolumn{3}{c|}{$\mathcal{P}$} & \multicolumn{3}{c|}{$\mathcal{P}_{cr}$} & \multicolumn{3}{c||}{$\mathcal{P}_{var}$} & \multicolumn{3}{c|}{$\mathcal{P}$} & \multicolumn{3}{c|}{$\mathcal{P}_{cr}$} & \multicolumn{3}{c|}{$\mathcal{P}_{var}$} \\ \hline
  $popk$ - \textit{ptb} & 1 & 2 & 3 & 1 & 2 & 3 & 1 & 2 & 3 & 1 & 2 & 3 & 1 & 2 & 3 & 1 & 2 & 3 & 1 & 2 & 3 & 1 & 2 & 3 & 1 & 2 & 3 \\ \hline
AUC & 0.5225 & \textbf{0.5234} & 0.5225 & 0.5000 & 0.5010 & 0.5003 & 0.5228 & 0.5225 & \underline{0.5230} & 0.5258 & \textbf{0.5500} & \underline{0.5316} & 0.5005 & 0.5096 & 0.5006 & 0.5007 & 0.5008 & 0.5013 & 0.4913 & 0.4902 & 0.4915 & 0.4989 & 0.4983 & \textbf{0.5000} & \underline{0.4991} & \underline{0.4991} & 0.4981\\ \hline
MRR & 0.2892 & \textbf{0.3133} & 0.2935 & 0.3004 & 0.2984 & 0.2636 & \underline{0.3070} & 0.2947 & 0.2863 & 0.3346 & \underline{0.3516} & 0.3274 & 0.3112 & \textbf{0.3760} & 0.3542 & 0.3356 & 0.3233 & 0.3245 & 0.2875 & 0.2845 & 0.3086 & 0.2767 & 0.2689 & \underline{0.3013} & 0.2796 & 0.2919 & \textbf{0.3180} \\ \hline
$nDCG@5$ & 0.1854 & \textbf{0.2117} & \underline{0.2058} & 0.1945 & 0.1953 & 0.1765 & \underline{0.2069} & 0.1921 & 0.1859 & 0.2419 & \underline{0.2570} & 0.2360 & 0.2214 & \textbf{0.2951} & 0.2512 & 0.2435 & 0.2353 & 0.2342 & 0.1939 & 0.1870 & \underline{0.2176} & 0.1948 & 0.1851 & 0.2073 & 0.1940 & 0.1973 & \textbf{0.2305} \\ \hline
$nDCG@10$ & 0.2708 & \textbf{0.3108} & \underline{0.2926} & 0.2867 & 0.3011 & 0.2648 & \textbf{0.3106} & 0.2829 & 0.2857 & 0.3415 & \underline{0.3511} & 0.3355 & 0.3165 & \textbf{0.3847} & 0.3339 & 0.3505 & 0.3342 & 0.3265 & 0.2908 & 0.2761 & \underline{0.3123} & 0.2937 & 0.2825 & 0.3071 & 0.3023 & 0.2961 & \textbf{0.3308} \\ \hline
\end{tabular}
}
\caption{Performance metrics using different popularity measures $\mathcal{P}(t, popk)$, $\mathcal{P}(t, popk){cr}$, and $\mathcal{P}(t, popk){var}$ with the \textit{POPK} logic $ptb$ and $popk$ values (1, 2, 3) on the \textit{Nikkei} dataset.}
\label{tab:results_ablation}
\end{table*}

\subsubsection{General Observations}
Precision in tailoring recommendations to match user preferences tends to narrow the range of recommended news types, while recommending a broader spectrum of articles typically reduces accuracy. Therefore, selecting a \textit{POPK} method requires careful consideration to balance accuracy and diversity according to specific objectives. It is crucial to note that this balance is nuanced and not straightforward, as evidenced by the chart in Figure \ref{fig:tradeoff}. The interplay among popularity, diversity, and accuracy is complex and influenced by user preferences. Each model and dataset may have an optimal configuration, depending on whether maximizing accuracy or enhancing diversity is prioritized. For instance, in the \textit{MIND-small} dataset, NAML performs strongly in category diversity ($D_{ctg}@5$) with $popk = 1 (\text{acc})$, although it shows weaker performance in nDCG@5. Conversely, in the \textit{Adressa one-week} dataset, LSTUR performs better for $popk = 3 (\text{acc})$ in terms of accuracy (nDCG@5) despite moderate category diversity ($D_{ctg}@5$).

\subsection{RQ4. Influence of Popularity Measurement}

We leveraged our \textit{Nikkei} dataset alongside corresponding baseline models to explore diverse paradigms for measuring popularity. Instead of focusing solely on \textit{number of clicks}, we evaluated popularity using two metrics: the highest click ratio at time \(t\), denoted as \(\mathcal{P}(t, popk)_{cr}\), and the greatest variation in clicks within time \(t\), denoted as \(\mathcal{P}(t, popk)_{var}\). Table \ref{tab:results_ablation} presents the performance metrics for each baseline model under different popularity measurement strategies, employing the $ptb$ logic and varying \(popk\) from 1 to 3. Our objective in this study is to illustrate how different methods of computing popularity metrics can yield distinct advantages and disadvantages. Additionally, we aim to emphasize the flexibility of defining different popularity functions for dynamic applications in industry settings. 

The results reveal that each model excels under specific measurements: NRMS achieves optimal performance with $\mathcal{P}(t, popk)$, NAML benefits most from $\mathcal{P}(t, popk)_{cr}$, and LSTUR demonstrates superior performance with $\mathcal{P}(t, popk)_{var}$. These findings emphasize the importance of customizing methods for computing popularity to improve the effectiveness of news recommendation systems across diverse scenarios. Therefore, different approaches to popularity computation can be tailored to achieve specific objectives in recommender systems, whether enhancing diversity or accuracy. Our study aimed to underscore the flexible adaptability of popular news retrieval functions, illustrating their ability to be customized for various domains and goals.

\section{RQ5. Limitations and Advantages}
We aim to delve into the limitations and advantages observed throughout our work, as every proposed model has its pros and cons. One significant drawback we noted is that, as previously discussed, the behavior of \textit{POPK} concerning accuracy and diversity can be intricate and not always straightforward, as evidenced by the \textit{Adressa one-week} dataset. Therefore, for unique datasets considering the application of the \textit{POPK} method, it may be necessary to conduct experiments to find the optimal balance between \textit{popk} values, logic (\textit{acc} and \textit{ptb}), and the most suitable popularity measurement function. 

Despite these challenges, we find that the \textit{POPK} method straightforwardly integrates into existing models, offering enhanced diversity without significant increases in computational complexity. Our experiments across various datasets indicate that the training time for models using \textit{POPK} remains comparable to baseline models. Unlike some graph algorithms that rely on pre-computed information from static global graphs, such as those mentioned by \cite{Yang_2023}, which restrict their applicability to dynamically changing real-world data, the \textit{POPK} method is designed to function with up-to-date and live data environments. In summary, while \textit{POPK} presents challenges in balancing accuracy and diversity, it offers a practical solution for improving recommendation systems without imposing prohibitive computational demands or reliance on static data structures.

\section{Conclusion}

This paper introduces \textit{POPK}, an novel approach employing temporal-counterfactual analysis to mitigate the influence of popular news articles. By examining a hypothetical scenario where a specified number \(popk\) of popular news articles competes with the user's preferred article at time \(t\), \textit{POPK} aims to enhance recommendation accuracy and diversity. We illustrate that our model effectively enhances accuracy and diversity within traditional recommender systems. Furthermore, we explore distinct computations of popularity, which varies according to different paradigms on how to define "popular" news articles. Extensive experiments on real-world datasets across Japanese, English, and Norwegian languages validate the efficacy of our approach. For future research, we are interested in exploring novel strategies for measuring popularity, particularly those that can adapt and learn over time.

\begin{acks}
This work is partially supported by "Joint Usage/Research Center for Interdisciplinary Large-scale Information Infrastructures (JHPCN)" in Japan (Project ID: jh241004), JSPS KAKENHI Grant Number 23K28098.
, and the Monbukagakusho: MEXT (Ministry of Education, Culture, Sports, Science and Technology - Japan) scholarship. 
\end{acks}


\bibliographystyle{ACM-Reference-Format}
\bibliography{popk}

\end{document}